\documentclass[usenatbib,psfig]{mn2e}

\usepackage{epsfig}

\begin{document}


\title[Bar ages and suppression of star formation]{Star formation suppression and bar ages in nearby barred galaxies.}
\author[P. A. James \& S. M. Percival]{P. A. James\thanks{E-mail:
P.A.James@ljmu.ac.uk} \& S. M. Percival\\
Astrophysics Research Institute, Liverpool John Moores University, IC2, Liverpool Science Park, 146 Brownlow Hill, Liverpool L3 5RF, UK\\
}

\date{Accepted . Received ; in original form }

\pagerange{\pageref{firstpage}--\pageref{lastpage}} \pubyear{2017}

\maketitle

\label{firstpage}

\begin{abstract}
 We present new spectroscopic data for 21 barred spiral galaxies, which we use to explore the effect of bars on disk star formation, and to place constraints on the characteristic lifetimes of bar episodes. The analysis centres on regions of heavily suppressed star formation activity, which we term `star formation deserts'.  Long-slit optical spectroscopy is used to determine  H$\beta$ absorption strengths in these desert regions, and comparisons with theoretical stellar population models are used to determine the time since the last significant star formation activity, and hence the ages of the bars.  We find typical ages of $\sim$1~Gyr, but with a broad range, much larger than would be expected from measurement errors alone, extending from $\sim$0.25~Gyr to $>$4~Gyr.  Low-level residual star formation, or mixing of stars from outside the `desert' regions, could result in a doubling of these age estimates. The relatively young ages of the underlying populations coupled with the strong limits on the current star formation rule out a gradual exponential decline in activity, and hence support our assumption of an abrupt truncation event. 

\end{abstract}


\section{Introduction}

Bars are a common feature in disk galaxies, with $\sim$30 per cent of galaxies having strong bars that are clearly apparent in their optical morphologies \citep{knap00,mari07,nair10}, $\sim$50 per cent of galaxies in the Sloan Digital Sky Survey (SDSS) showing bars \citep{bara08,ague09}, and at least 60 per cent showing evidence of stellar bars in near-infrared imaging \citep{seig98,gros04,mene07}.  There has been extensive discussion of the possible central role of bars in the so-called `secular evolution' of disk galaxies \citep{korm04}, with the main debate centring on the possibility that bars build central bulge components at the expense of disk material \citep{pfen90,frie93,cheu13}.  In this picture, bars are a major factor in constructing the morphological sequence of changing bulge-to-disk ratio first pointed out by \cite{hubb26}. Recently, an additional aspect of this debate has emerged, with the suggestion that bars can play a controlling aspect in the overall star formation (SF) rate of disk galaxies, and may indeed be a significant player in the transformation of galaxies between the `blue cloud' and  the `red sequence' populations first identified by \cite{stra01} and \cite{bald04}.   The possibility that bars may decrease total SF rates within a galaxy was proposed by \cite{tubb82} and has seen renewed interest recently in the proposal of \cite{hayw16} that a turn-down in the SF rate of the Milky Way may have been driven by the simultaneous formation of the Galactic bar.  This suggestion has also received confirmation in the context of cosmological studies of large samples of galaxies at low and intermediate redshifts; \cite{gava15} and \cite{cons17} find that that the SF rate vs. stellar mass relation of strongly barred galaxies is offset significantly towards lower SF values, compared to unbarred galaxies. Similarly, \cite{kim17} use a large sample of galaxies from the Sloan Digital Sky Survey to demonstrate that the strongly-barred galaxies show significantly lower SF activity than their unbarred counterparts.

Our work in this area started with the analysis of the spatial distributions of SF in the disks of barred and unbarred galaxies using narrow-band H$\alpha$ imaging from the H$\alpha$GS survey \citep{jame04}. In \cite{jame09} this imaging was used to demonstrate that strongly-barred galaxies, particularly of earlier Hubble types, exhibit a radial pattern of SF that is radically different to that of their unbarred counterparts.  The defining feature within the barred galaxies is the almost complete suppression of SF in the radial range `swept out' by the bar, which we term the `star formation desert' (SFD).  This is accompanied in many cases by an {\em increase} in SF both at the centre of the bar and in a ring just outside the ends of the bar, with the bar-end increase generally being larger than that found in nuclei, in terms of emission line flux and inferred SF activity. 

Strong supporting evidence for this SFD phenomenon was provided from a completely independent line of analysis by \cite{hako16}.  They analysed the radial distributions of $\sim$500 core-collapse supernovae in nearby disk galaxies, which showed clear suppression of SN numbers in the radial ranges swept out by strong bars, just as predicted by the SFD hypothesis.  Interestingly, this suppression effect was only found in Sa to Sbc barred galaxies, and not those of type Sc and later, consistent with our findings from H$\alpha$ profiles in \cite{jame09}, and equivalent findings of `bar bimodality' by \cite{nair10} and \cite{hako14} \citep[see also the theoretical predictions of][]{vill10}, with late-type bars being associated with concentrated SF along the length of bars, rather than deserts.  Further confirmation came from the recent study of \cite{abdu17}, who found lower specific SF rates in the central regions, but not in the outer regions, for the barred galaxies within a sample of 93 low-redshift massive spiral galaxies.

A possible mechanism for the SF suppression effect of bars was identified by \cite{reyn98}.  In a detailed study of the nearby strongly-barred SBb galaxy NGC\,1530, they found that the bar induces strong shocks and shearing gas motions which they conclude should be sufficiently strong to stabilise the gas against collapse and subsequent SF.  Similarly, \cite{verl07} concluded that gas velocities induced by bars may be sufficient to stabilise even high gas densities against SF in the immediate vicinity of the bar.  An alternative possibility is suggested by the study of \cite{spin17}.  They simulate the evolution of a Milky Way-like spiral galaxy, of estimated type Sb or Sc, and find that the formation of a bar efficiently depletes gas in a region of radius $\sim$2~kpc, corresponding to the maximum radial length of the bar.  Deciding between these possibilities of stabilisation or complete removal of gas from SFD regions is beyond the scope of the present paper, but is an interesting question for future study.

\subsection{Stellar populations in star formation desert regions}

In \cite{jame16} we presented our first detailed study of stellar populations within  SFD-hosting galaxies, using long-slit optical spectroscopy of four strongly-barred early-type galaxies, NGC\,2543, NGC\,2712, NGC\,3185 and NGC\,3351. This study confirmed the complete suppression of SF within the SFD region in these 4 galaxies; while some low-level diffuse line emission was detected, this showed line ratios and a smooth spatial distribution that can be much more naturally explained by an old stellar population, e.g. post-AGB stars \citep{jame15}; see also \cite{stas08}, \cite{sarz10}, \cite{yan12} and \cite{brem13}. After correcting for this weak line emission, we fitted the remaining absorption-line spectra by a model that assumed an initially constant SF rate, followed by a sharp truncation in SF activity; the one free parameter in the model was the epoch at which this truncation occurred. The most important conclusion of this analysis was that SF was suppressed long ago in the SFD regions of these four galaxies, and that this epoch differs significantly between the four, with a range of $\sim$1 to $>$4~Gyr.

The present paper builds on the work presented by \cite{jame16} by substantially increasing the sample size of observed galaxies.  New long-slit spectroscopic data are presented and analysed for a total of 21 barred galaxies, comprising repeated observations for 3 of the galaxies from \cite{jame16} and new observations for 18 previously unstudied systems.

\section{Observations}

\begin{figure}
\includegraphics[width=90mm,angle=0]{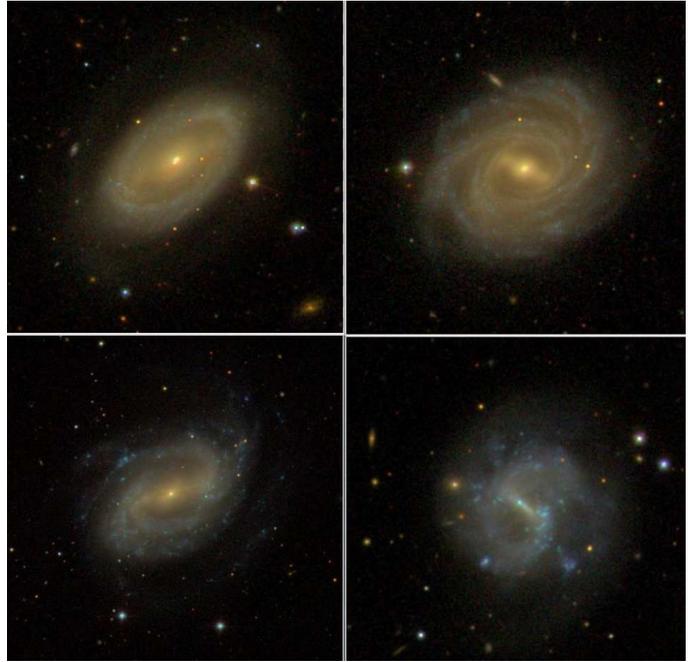}
\caption{
SDSS $gri$ images of four of the observed barred galaxies, showing the range of Hubble type classifications of the full sample.  
Top left: NGC\,3185, classification (R)SB(r)a.
Top right: NGC\,4999, classification SB(r)b.
Bottom left: NGC\,4123, classification SB(r)c, for which the H$\beta$ absorption strength could only be extracted for one side of the SFD.
Bottom right: NGC\,2604, classification SB(rs)cd, which shows strong SF at all locations, including within the bar and in the regions swept out by the bar, and hence is not included in the main analysis, as explained in Sect. 3.1.
}
\label{fig:barsx4}
\end{figure}

The properties of the 21 barred galaxies in the present study are listed in Table~\ref{tab:gals_obs}, including classification and distance properties which were taken from the NASA/IPAC Extragalactic Database (NED).  Column 1 of Table~\ref{tab:gals_obs} contains the galaxy name; column 2 the galaxy classification; column 3 the heliocentric recession velocity; column 4 the adopted galaxy distance, based on a Hubble constant of 73~kms$^{-1}$Mpc$^{-1}$ and a correction for infall velocities generated by the Virgo cluster; and column 5 the adopted bar position angle. Column 6 gives the position angle of the slit in the new spectroscopic observations presented in this paper and column 7 the total integration time in seconds for these observations. Finally, column 8 gives the source of the imaging observations used for initial classification and for the measurement of bar parameters.  An entry of `H$\alpha$' in this column means that our own H$\alpha$ imaging, from \cite{jame04}, was used to identify the galaxy as possessing a SFD, and to measure the radial extent of the SFD region when extracting integrated spectra.  $R$-band images from the same source were used to determine the bar position angle.  An entry of `SDSS' in column 8 means that Sloan Digital Sky Survey \cite{york00} $gri$ colour images \citep{lupt04} were used for all these tasks.

In comparison with our previous work \citep{jame16}, we are now able to present not only a much larger sample, but also higher quality and more uniform spectroscopic data, with our earlier observations having been seriously affected by poor observing conditions.  
Galaxies in the new sample lie at somewhat greater distances (range 14 - 80~kpc, cf. 10 - 40~kpc for the earlier study), and cover a wider range of Hubble type.  In \cite{jame16}, one galaxy was of type Sa and the other three Sb, and all were strongly barred with SB classifications. In this study, we present results for 2 Sa, 1 Sab, 9 Sb, 2 Sbc, 5 Sc and 2 Scd galaxies. Of these 21 galaxies, 18 have strongly-barred SB classifications, and the other 3 are considered intermediate-strength SAB types.  We checked whether any galaxies within this sample are known to host a double bar (otherwise known as a `bar-within-a-bar') by cross-checking with the latest version of the list maintained by Peter Erwin, first published as \cite{erwi04}.  Possibly surprisingly, only NGC\,5806 is a known double-bar host, which may indicate that the SFD phenomenon is less common or less marked in galaxies with these central components.

The new observations presented here are long-slit optical spectra,
taken with the Intermediate Dispersion Spectrograph (IDS) on the 2.5-metre Isaac Newton Telescope (INT) at the
Observatorio del Roque de Los Muchachos on La Palma in the Canary
Islands.  The observing time was allocated to proposal I/2016/P1 by the UK Panel for the Allocation of 
Telescope Time. The observing dates allocated to the project were the 7 nights of the 3rd to 9th March 2016 inclusive, with usable data being taken on all nights. Occasional thin cirrus cloud affected 3 nights, and the first two nights suffered from haze due to Saharan dust in the atmosphere above the observatory; however, photometric conditions were not required for these observations.

The observing set-up of the IDS was as follows. The grating used was R900V, which has a blaze wavelength of 5100\,\AA. The EEV10 blue-sensitive CCD detector was used to maximise sensitivity in the region of H$\beta$ absorption, the most important feature for the present investigation. The slit width for all galaxy observations was 1.5$^{\prime\prime}$, corresponding to 2.9 pixels or 1.8\,\AA\ in the spectral direction, given a dispersion of 0.63\,\AA~pixel$^{-1}$.  The spatial resolution for this setup is 0.4$^{\prime\prime}$/pixel, and an unbinned CCD readout was used.  The unvignetted wavelength coverage was 1518\,\AA\ about a central wavelength of 4900\,\AA. The slit was generally aligned to be perpendicular or close to perpendicular to the bar of each galaxy (see Table~\ref{tab:gals_obs}) and so in general observations were not made at the parallactic angle. However, all of the important parameters for the present analysis are derived from line equivalent width (EW) values which are unaffected by atmospheric dispersion effects.

Spectrophotometric standard star observations, using sources selected from the observatory standards list\footnote{http://catserver.ing.iac.es/landscape/tn065-100/workflux.php} were made 2 - 3 times per night through a wide slit (4 - 5$^{\prime\prime}$) to provide flux calibration.  However, this calibration was not required for any of the results presented here.

\begin{table*}
 \begin{minipage}{140mm}
  \caption{Galaxy properties and observing details.
See Section 2 for descriptions of column entries.}
  \begin{tabular}{llccrrrc}
  \hline
 Name     &    Classn      &   Vel          & Dist &  Bar PA   &   Slit PA  &  T$_{int,tot}$ & SFD Image\\
          &                & (km\,s$^{-1}$)  & (Mpc) &  (Deg)     &   (Deg)  &  (sec)         \\
  \hline
UGC\,3685   &   SB(rs)b     &   1797   &   28.3    &   131     &       0    &  10800   & H$\alpha$ \\
NGC\,2487   &   SB(r)c      &   4827   &   66.8    &    45     &     315    &  10800   & SDSS \\
NGC\,2523   &   SB(r)bc     &   3471   &   51.5    &   125     &      60    &   8400   & SDSS \\
NGC\,2543   &   SB(s)b      &   2470   &   35.9    &    90     &     180    &   7200   & H$\alpha$ \\
NGC\,2595   &   SAB(rs)c    &   4324   &   59.9    &   158     &     260    &   7200   & H$\alpha$ \\
NGC\,2712   &   SB(r)b      &   1815   &   28.1    &    28     &     350    &   6000   & H$\alpha$ \\
NGC\,3185   &   (R)SB(r)a   &   1238   &   20.0    &   113     &     320    &   7200   & H$\alpha$ \\
NGC\,3346   &   SB(rs)cd    &   1250   &   20.4    &    95     &       0    &  13200   & SDSS \\
NGC\,3367   &   SB(rs)c     &   2944   &   42.1    &    72     &     170    &   7200   & H$\alpha$ \\
NGC\,3485   &   SB(r)b      &   1436   &   24.0    &    40     &     130    &   7200   & SDSS \\
NGC\,3507   &   SB(s)b      &    979   &   13.8    &   120     &      30    &  13200   & SDSS \\
NGC\,3729   &   SB(r)a      &   1060   &   19.4    &    20     &     290    &   6000   & SDSS \\
NGC\,3811   &   SB(r)cd     &   3105   &   46.4    &    30     &     267    &   6000   & H$\alpha$ \\
NGC\,4123   &   SB(r)c      &   1286   &   25.3    &   105     &      15    &   7200   & SDSS \\
NGC\,4779   &   SB(rs)bc    &   2855   &   41.9    &     5     &     280    &   7200   & SDSS \\
NGC\,4999   &   SB(r)b      &   5634   &   78.7    &    60     &     330    &   6000   & SDSS \\
NGC\,5350   &   SB(r)b      &   2311   &   36.5    &   130     &      20    &   6000   & SDSS \\
NGC\,5375   &   SB(r)ab     &   2373   &   37.0    &   170     &     260    &   6000   & SDSS \\
NGC\,5698   &   SBb         &   3629   &   54.1    &   168     &      70    &   6000   & H$\alpha$ \\
NGC\,5806   &   SAB(s)b     &   1347   &   22.4    &   172     &      30    &   8400   & H$\alpha$ \\
NGC\,5970   &   SB(r)c      &   1957   &   30.9    &    80     &     350    &   7200   & SDSS \\
\hline
\end{tabular}
\label{tab:gals_obs}
\end{minipage} 
\end{table*}

\section{Data reduction}

\subsection{Basic reduction and spectral extraction}

Data reduction of our long-slit spectroscopy was performed using Starlink software \citep{curr14}.  All stages, i.e. bias subtraction, flat fielding, correction of spectra for minor rotation and spatial distortion effects, sky background subtraction, wavelength calibration using a copper neon $+$ copper argon arc lamp and flux calibration from spectrophotometric standard observations were completely standard are will not be described further here.  More details are given in \cite{jame16}.  H$\alpha$ narrow-band imaging, where available (see Table~\ref{tab:gals_obs}), was used to define the apparently emission-line free region from which the `desert' spectra were extracted. In other cases, SDSS $gri$ colour images were used for this purpose. In all cases, two desert regions were extracted for each galaxy, one on either side of the nucleus, and the radial ranges were conservatively restricted to ensure that all emission from the nuclear regions was excluded.
 
The new analysis presented here resulted in 42 spectra, one from each side of the nucleus of each of the 21 galaxies.  As noted above, this sample includes 3 of the 4 galaxies discussed in \cite{jame16}, but we have obtained new, deeper spectroscopy for these galaxies, and the updated measurements are used here.

The spectral analysis methods used were described in full in \cite{jame16}.  The main parameters extracted from the spectra were absorption line EW values.  The most important feature, due to its strong age sensitivity, is H$\beta$ 4861\,\AA, but in order to take account of age-metallicity degeneracy in the strength of this feature, we also measured the predominantly metallicity sensitive Mgb feature centred on 5176\,\AA.  EW values for these features were determined by direct integration on the reduced and extracted spectra, using the wavelength ranges for features and continuum defined by \cite{trag98}. 

Rotational velocities and stellar velocity dispersion could in principle have an effect on measured line indices.  For the present sample, galaxy rotation will not have a large effect; the galaxies observed are generally close to face-on, and the spectra are extracted separately from opposite sides of the nucleus so the rotational difference between these regions is not an issue.  Departure from flat rotation curves in the extracted regions, and velocity dispersion, will lead to moderate broadening of absorption features, but this will at most be at the 100\,km\,s$^{-1}$ level.  We conservatively tested the effects of this by convolving BaSTI solar-abundance models with Gaussian functions with a $\sigma$ of 150 km\,s$^{-1}$, and found a completely negligible impact on the indices used here, which is understandable given that the indices used are equivalent widths summed over broad bandpasses (29\,\AA\ for the most important index, H$\beta$, while 150\,km\,s$^{-1}$ is equivalent to only 2.4\,\AA\ at the wavelength of H$\beta$).

There is one case of a galaxy that was observed in the March 2016 observing run, but where no results are presented in this paper.  This was the galaxy NGC\,2604, of type SB(rs)cd.  In this case, strong emission lines with ratios consistent with expectations for SF regions \citep{bald81} were observed at all locations along the slit, so there is nothing that can be termed a SFD in this galaxy, and these lines are sufficiently strong to prevent the accurate measurement of absorption line indices, both at H$\alpha$ and H$\beta$.  This is not entirely surprising; NGC\,2604 is described as an `Interacting galaxy with deformed arms' by \cite{miya98} in their study of ultra-violet excess galaxies, and the NED classification notes Wolf-Rayet and starburst characteristics.  This galaxy also stands out as unusually blue in the SDSS $gri$ colour composite image (see Fig.~\ref{fig:barsx4}, lower-right frame); the blue colours extend over the whole of the galaxy, including the bar, so this is clearly an example where a bar has not suppressed SF anywhere in the disk or bar regions.

\subsection{Correction for line emission}

In \cite{jame15} we used long-slit spectroscopy in the red part of the optical spectrum around the H$\alpha$ emission line to show that almost all SFD regions in the 15 spiral galaxies studied exhibit low-level diffuse line emission.  However, the H$\alpha$/[NII] line ratios are characteristic of LINER-type regions \citep{bald81,heck80a} which conclusively excludes SF as the powering source behind this emission.  We have extended that analysis to encompass a much larger sample of galaxies, including all 21 of the barred galaxies that are the subject of this paper.  The analysis of the emission-line properties of the enlarged sample will be presented in a separate paper (James \& Percival, in preparation). Here it need only be noted that 9 of the present sample (those with an entry of `H$\alpha$' in column 8 of Table~\ref{tab:gals_obs}) have emission line fluxes already published in \cite{jame15} while the remaining 12 have emission-line properties that will be discussed in the future paper.  There is no significant difference in the data quality or analysis methods between the two sub-samples.

As with \cite{jame16}, the emission-line measurements are essential for the accurate correction of emission-line contamination of the H$\beta$ absorption features.  Without such correction, the H$\beta$ strengths would be biased to significantly lower values, resulting in significant over-estimates of stellar population ages.  To avoid this, we used the same iterative emission line correction procedure as in \cite{jame16}, which is based on the assumption of Case B line ratios for Balmer emission lines \citep{broc71,humm87}.  The second constraining assumption of the correction method is that the relative strengths of the H$\alpha$ and H$\beta$ absorption features should be consistent with the predictions of theoretical stellar population models \citep{piet04,perc09}.  In the implementation of this correction, the H$\beta$ absorption EW was measured first; this was then used to predict an H$\alpha$ absorption EW based on the stellar population model; this absorption strength was used to give a corrected H$\alpha$ emission line flux, and hence an H$\beta$ line flux following the Case B assumption.  The latter value was used to correct the initially measured H$\beta$ absorption line EW, which was then taken as the starting point for the second iterative loop.  In practice, the values converged quickly to values that satisfied the requirements of fitting the population models and expectations of the Case B assumption, after 3 or 4 iterations. 

The results of the data reduction processes are shown in Table~\ref{tab:reg_specprops}.  Here the first column gives the galaxy name, and the second column gives an identifier for the region from which the integrated SFD spectrum was taken.  Column 3 gives the radial range in arcseconds of the region used to extract the spectrum, and column 4 the direction of offset of this region from the galaxy nucleus.  Column 5 gives the initial measurement of the H$\beta$ absorption-line EW in \AA, with the associated error in column 6, and column 7 the H$\beta$ EW after the iterative emission-line correction described in the previous paragraph.

Three of the SFD regions listed in Table~\ref{tab:reg_specprops} do not have values for the emission-line corrected H$\beta$ EW.  These regions lie in NGC\,2487, NGC\,2595 and NGC\,4123, all of which have Hubble types of SBc (actually SABc for NGC\,2595), and hence three of the latest types represented in the present sample.  Thus it is not surprising that these should have some residual SF in their central disk regions. One of these galaxies, NGC\,4123, is shown in Fig.~\ref{fig:barsx4} (lower-left frame), which confirms that is indeed a transitional case between the `true' SFD galaxies with red central disks shown in the upper row, and the starburst galaxy NGC\,2604 to its right. The residual SF in the SFD regions of NGC\,2487, NGC\,2595 and NGC\,4123 then explains the strong H$\beta$ emission that prevents the measurement of the absorption feature in one half of the disk. The other half of the extracted SFD region in these three cases does give an acceptable iterative solution, generally resulting in a strong underlying absorption, indicating recent truncation ages and/or contamination by a younger stellar component. 

The importance of accurate emission-line correction can be seen by comparing columns 5 and 7 of Table~\ref{tab:reg_specprops}, with the H$\beta$ EW values increasing following this correction by several times the measurement error in most cases. Indeed, about one-third of the regions would have unphysically small H$\beta$ EW values without this correction, i.e. lower than could be fitted by any stellar population younger than the age of the Universe. The importance of this correction is not apparent from inspection of the spectra; the only ones that show any evidence for an emission peak at H$\beta$ are those extracted from NGC\,2595 and NGC\,4123 with negative EW values in column 5 of Table~\ref{tab:reg_specprops}.  Otherwise, the emission is lost in the core of the absorption line profile.

\begin{figure}
\includegraphics[width=90mm,angle=0]{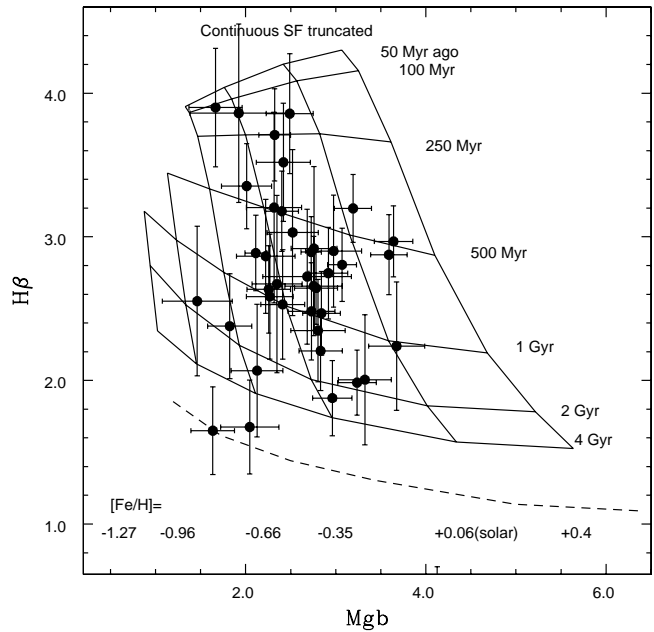}
\caption{Observed emission-corrected H$\beta$ and Mgb absorption strengths for the 39 SFD regions with well-measured H$\beta$ EW values (see Table~\ref{tab:reg_specprops}), overlaid on a grid of model predictions for different scaled-solar metallicities and elapsed times since cessation of SF.  The grid is derived from models with a spectral resolution of 1.0\,\AA, while the observations have a resolution of 1.83\,\AA.  The observed range of SF truncation ages is from 0.1 to at least 4~Gyr, much larger than the expected range from uncertainties in age determinations.  Even given this broad range of ages there is some indication of preferred timescale, with points clustering around an age of $\sim$1~Gyr.
}
\label{fig:sfdgrid_e}
\end{figure}

\begin{table*}
 \begin{minipage}{140mm}
  \caption{Observed regions and derived spectroscopic properties.}
  \begin{tabular}{ccccccc}
  \hline
Galaxy   &  Region  & Radial range  & Dir  &   H$\beta_{obs}$ &  err &   H$\beta_{corr}$  \\
         &          & ($^{\prime\prime}$) & &     (\AA)      &       &      (\AA)  \\
\hline
UGC\,3685  &   SFD1   &   7.48-17.60   &    N    &   0.877   &    0.237   &    1.650   \\
UGC\,3685  &   SFD2   &   7.48-17.60   &    S    &   1.255   &    0.234   &    2.378   \\
NGC\,2487  &   SFD1   &   6.16-19.36   &    SE   &   1.558   &    0.248   &    2.481   \\
NGC\,2487  &   SFD2   &   6.60-14.08   &    NW   &   0.285   &    0.311   &    -----   \\
NGC\,2523  &   SFD1   &   6.20-26.40   &    SW   &   1.559   &    0.199   &    1.985   \\
NGC\,2523  &   SFD2   &   6.20-21.60   &    NE   &   2.425   &    0.206   &    2.968   \\
NGC\,2543  &   SFD1   &   7.48-24.20   &    N    &   2.646   &    0.269   &    3.900   \\
NGC\,2543  &   SFD2   &   6.60-19.80   &    S    &   2.500   &    0.243   &    3.858   \\
NGC\,2595  &   SFD1   &   9.68-19.36   &    E    &  --0.276  &    0.568   &    -----   \\
NGC\,2595  &   SFD2   &  13.20-26.40   &    W    &   2.335   &    0.491   &    3.861   \\
NGC\,2712  &   SFD1   &   6.16-21.56   &    S    &   1.776   &    0.207   &    2.747   \\
NGC\,2712  &   SFD2   &   7.48-20.68   &    N    &   1.323   &    0.233   &    2.529   \\
NGC\,3185  &   SFD1   &   9.68-21.56   &    SE   &   1.575   &    0.227   &    2.206   \\
NGC\,3185  &   SFD2   &  11.44-29.04   &    NW   &   1.257   &    0.211   &    1.877   \\
NGC\,3346  &   SFD1   &   6.60-10.56   &    N    &   1.050   &    0.361   &    2.552   \\
NGC\,3346  &   SFD2   &   5.28-11.44   &    S    &   0.630   &    0.289   &    2.068   \\
NGC\,3367  &   SFD1   &   5.28-9.68    &    N    &   0.905   &    0.298   &    2.239   \\
NGC\,3367  &   SFD2   &   5.72-10.12   &    S    &   0.605   &    0.287   &    2.005   \\
NGC\,3485  &   SFD1   &   5.28-14.52   &    NW   &   1.288   &    0.312   &    1.675   \\
NGC\,3485  &   SFD2   &   5.28-12.76   &    SE   &   2.759   &    0.257   &    3.353   \\
NGC\,3507  &   SFD1   &   6.16-23.76   &    NE   &   1.932   &    0.111   &    2.887   \\
NGC\,3507  &   SFD2   &   6.16-26.40   &    SW   &   1.997   &    0.105   &    2.894   \\
NGC\,3729  &   SFD1   &   4.40-14.10   &    NW   &   0.458   &    0.272   &    2.671   \\
NGC\,3729  &   SFD2   &   6.20-14,50   &    SE   &   0.873   &    0.250   &    2.919   \\
NGC\,3811  &   SFD1   &   5.28-10.56   &    W    &   1.915   &    0.302   &    2.901   \\
NGC\,3811  &   SFD2   &   5.28-10.56   &    E    &   2.036   &    0.308   &    2.657   \\
NGC\,4123  &   SFD1   &   8.80-26.40   &    N    &   0.811   &    0.287   &    3.204   \\
NGC\,4123  &   SFD2   &   8.80-23.32   &    S    &  --1.345  &    0.374   &    -----   \\
NGC\,4779  &   SFD1   &   5.28-13.64   &    E    &   1.000   &    0.276   &    3.030   \\
NGC\,4779  &   SFD2   &   5.28-18.04   &    W    &   1.141   &    0.243   &    2.584   \\
NGC\,4999  &   SFD1   &   6.16-16.30   &    NW   &   1.777   &    0.289   &    2.865   \\
NGC\,4999  &   SFD2   &   6.16-20.68   &    SE   &   2.266   &    0.267   &    3.519   \\
NGC\,5350  &   SFD1   &   6.16-13.64   &    N    &   1.825   &    0.227   &    2.634   \\
NGC\,5350  &   SFD2   &   6.60-16.72   &    S    &   2.141   &    0.220   &    2.643   \\
NGC\,5375  &   SFD1   &   4.40-19.36   &    W    &   2.086   &    0.197   &    2.875   \\
NGC\,5375  &   SFD2   &   4.40-20.24   &    E    &   2.625   &    0.188   &    3.198   \\
NGC\,5698  &   SFD1   &   6.16-22.00   &    SW   &   1.540   &    0.294   &    2.347   \\
NGC\,5698  &   SFD2   &   7.04-19.36   &    NE   &   2.229   &    0.453   &    2.723   \\
NGC\,5806  &   SFD1   &   6.60-13.20   &    SW   &   1.989   &    0.152   &    2.805   \\
NGC\,5806  &   SFD2   &   6.60-11.44   &    NE   &   1.640   &    0.206   &    2.467   \\
NGC\,5970  &   SFD1   &   4.40-8.80    &    N    &   2.295   &    0.172   &    3.178   \\
NGC\,5970  &   SFD2   &   4.40-9.68    &    S    &   2.605   &    0.164   &    3.709   \\
\hline
\end{tabular}
\label{tab:reg_specprops}
\end{minipage} 
\end{table*}

\section{Results}

\subsection{Star formation truncation ages from H$\beta$ absorption strengths}

Figure~\ref{fig:sfdgrid_e} shows the main results from this paper.  Each plotted point corresponds to the measured H$\beta$ absorption EW value from an individual SFD region, after correction for line emission, plotted against the metallicity-sensitive Mgb EW values, derived from direct integration on the observed SFD spectra using the wavelength ranges specified by \cite{trag98}.  Error bars are derived from the photometric errors on this direct integration process, plus an additional error added in quadrature for the H$\beta$ index only, to account for the additional uncertainty due to the emission-line correction. The points are overlaid on a grid derived from the BaSTI stellar population synthesis models \cite{piet04, perc09}.  As in \cite{jame16}, these theoretical grids are derived from a model where SF began 13~Gyr ago, and proceeded at a constant rate until an abrupt and complete cessation of SF at a later epoch, which we allow to vary from 4~Gyr ago (i.e. after 9~Gyr of steady SF) to only 50~Myr ago.  The calculation of these models for a range of scaled solar metallicities from [Fe/H] of --1.27 to +0.4 (strongly sub-solar to significantly above solar) results in the grid points plotted in Fig.~\ref{fig:sfdgrid_e}.  The dashed line at the bottom of that figure is the locus of values for varying metallicity but a single population age of 14~Gyr, and effectively represents the physical limit for how weak the H$\beta$ feature can become.

The main conclusions to draw from Fig.~\ref{fig:sfdgrid_e} are the following.  Almost all of the points lie on the model grid, and all have physically plausible values of H$\beta$ EW, i.e. they lie above the dashed line.  The inferred truncation ages, i.e. the time since the last strong episode of SF cover virtually the full range of modelled values, with two regions lying above the oldest modelled age of 4~Gyr (although consistent with this value given the measurement error bars), while the most recently truncated region lies on the 100~Myr line.  The spread in ages appears real, i.e. much greater than would be expected from measurement errors alone.

Figure~\ref{fig:sfdgrid_m} shows the same information as Fig.~\ref{fig:sfdgrid_e}, i.e. H$\beta$ and Mgb line indices overlaid on the BaSTI model grid, but presented on a galaxy-by-galaxy basis.  In this figure, index locations for pairs of regions that lie within the same galaxy are joined by dashed lines, with open circles representing the mean location in the index--index plane of each such pair.  Eighteen galaxies had well-measured spectroscopic properties in SFD regions on both sides of the galaxy nucleus, and dashed lines are shown for all of these in Fig.~\ref{fig:sfdgrid_m}. For the 3 galaxies previously mentioned, only one SFD spectrum could be reliably fitted to give absorption index determinations; in Fig.~\ref{fig:sfdgrid_m} these are plotted as circles with no associated dashed line. 

The data plotted in Fig.~\ref{fig:sfdgrid_m} enable us to test whether there are differences in stellar population properties for spatially-separated parts of the SFD regions {\em within} a given galaxy.  The generally short dashed lines in this figure indicate that any such inhomogeneities must be small, but we were able to quantify this impression by analysing the differences in observed corrected H$\beta$ absorption strengths in the 18 galaxies.  The median difference in EW was 0.334~\AA, while the median error on the EW measurements for the same 18 galaxies (36 regions) was 0.336~\AA.  Thus, the differences seen within galaxies can be completely explained by the effects of measurement errors, and there is no evidence for any inhomogeneities in spectroscopic properties within the SFD regions of these 18 galaxies. 

\subsection{Analysis of SFD truncation ages as a function of galaxy type}

In Fig.~\ref{fig:sfdgrid_m}, galaxy classifications are indicated by different point types.  There are no striking trends, but some inferences can still be drawn.  Most of the main peak of galaxies within this plot, at inferred bar ages of 0.5 - 1~Gyr, are of type SBb, the classification that we have found to be most strongly associated with the SFD phenomenon.  There is a slight excess of late-type galaxies with young bar ages, and/or larger fractional contamination of the investigated disk regions by young stars.  Two of the three youngest, and three of the five youngest, have classifications of Sc or Scd. In addition, it should be remembered that there are three regions in Sc galaxies for which no absorption measurement was possible due to strong emission lines, and another late-type, NGC\,2604, shows strong SF in all locations.  So there is some trend for older populations, and earlier truncation epochs, in early-type galaxies.  However, the main conclusion is that a broad range of truncation epochs, and hence inferred bar ages, is seen for all galaxy types. {NGC\,5806, the one galaxy identified as possessing a double bar, has entirely typical properties for a galaxy of its type, with a truncation age of just under 1\,Gyr.

\subsection{Effects of residual star formation or stellar mixing}

As an alternative to the models corresponding to complete truncation of SF at a given epoch, we also explored the effect of a downward step in SF, after which SF continued but only at 10 per cent of its previous rate.  This was motivated by the findings of the Milky Way SF rate study of \cite{hayw16}, who found just such a 90 per cent reduction in the SF rate which they ascribe to the development of the strong Galactic bar.  This obviously results in a larger fraction of young stars being present than for the earlier, complete truncation model, and so the epoch of such a step has to be pushed to earlier times to give an overall population that fits a given strength of H$\beta$ absorption.  To a fair approximation, we found that these `step ages' are about twice the `truncation ages' represented in the grid shown in Figs.~\ref{fig:sfdgrid_e} - \ref{fig:sfdgrid_m}, at least for the ages relevant to the bulk of the galaxy regions investigated here.  Thus, complete truncation at 0.5~Gyr resembles a 90 per cent step at 1~Gyr, 1~Gyr matches 2~Gyr, and 2~Gyr matches 4~Gyr.  Given the uncertainty of the appropriate step size to assume, we do not present full grids for this case, but the general results are instructive.

Note that many of our SFD regions show reductions in SF rate that are much more extreme than this 90 per cent value, as indicated by the weakness of any detected emission lines and the fact that any emission lines that are detected do not have ratios characteristic of SF. However, even if SF were to be completely suppressed within the SFD regions, as seems plausible, it is possible that young stars from elsewhere in the galaxy will be mixed in to the SFD region, e.g. through bar-driven tidal torques.  This would have the effect of introducing a younger stellar population without requiring ongoing SF within the region itself, and our 90 per cent step model is also useful to indicate the likely effects of such mixing on inferred bar ages.

\begin{figure}
\includegraphics[width=90mm,angle=0]{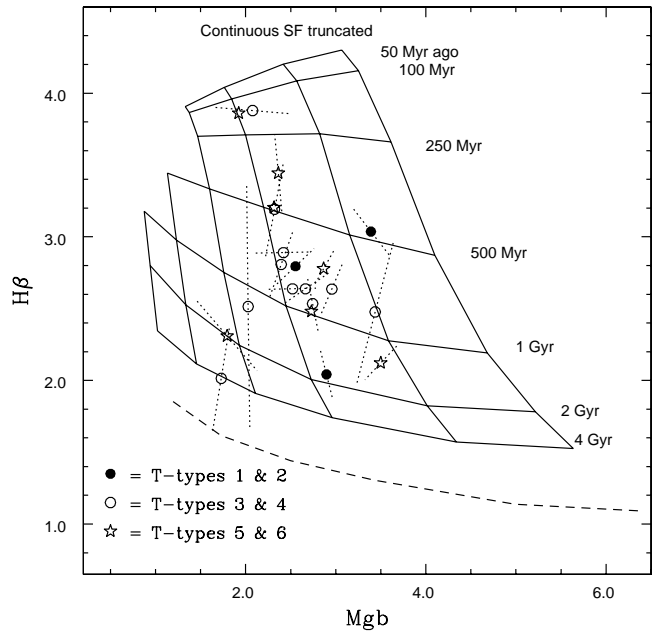}
\caption{As Fig.~\ref{fig:sfdgrid_e}, but now observations from SFD regions within the same galaxy are joined by dashed lines, with an open circle indicating the mean value of the derived age- and metallicity-sensitive parameters for that galaxy.  Circles with no associated lines correspond to galaxies with only one measured SFD region.  Different point types indicate the T-type classifications of the galaxies.
}
\label{fig:sfdgrid_m}
\end{figure}

\subsection{Exponentially-declining star formation histories} 

In Fig.~\ref{fig:sfd_exp_grid}, the SFD points from Fig.~\ref{fig:sfdgrid_m} are overplotted on a BaSTI model grid for an assumed exponentially-declining SF history.  For each grid point, SF is assumed to start 13.5\,Gyr ago, and the different models are distinguished by exponential time constant $\tau$ and metallicity.  Because of the lack of young, low-metallicity models in BaSTI, these models are now restricted to just four metallicity bins.  The SFD stellar populations cluster around models with exponential decay times of $\sim$6\,Gyr which has significant implications for the main analysis presented here.  If the SFD regions really had SF histories with these exponential properties, the SF rate would be predicted still to be on the order of 10 per cent of the initial value, as the galaxy ages are only about two exponential decay times.  Current SF at this rate is clearly ruled out by the weak emission lines seen as the defining feature of the SFD regions.  Thus, the combination of relatively strong H$\beta$ absorptions and stringent constraints on emission line fluxes requires there to have been a specific SF truncation event in these regions.

\begin{figure}
\includegraphics[width=90mm,angle=0]{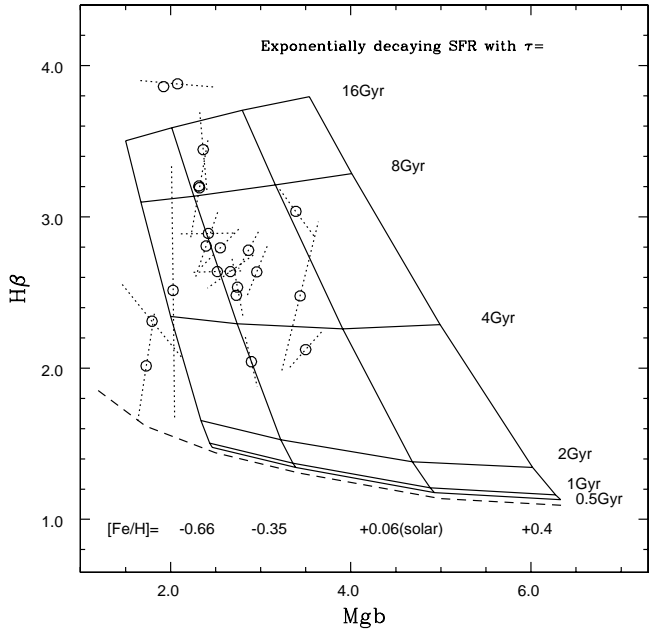}
\caption{The line index points from Fig.~\ref{fig:sfdgrid_m} are here overlaid on a grid of index values for models with exponentially-declining SF rates, starting 13.5\,Gyr ago, with time constants indicated to the right of the grid.}
\label{fig:sfd_exp_grid}
\end{figure}

\begin{figure}
\includegraphics[width=90mm,angle=0]{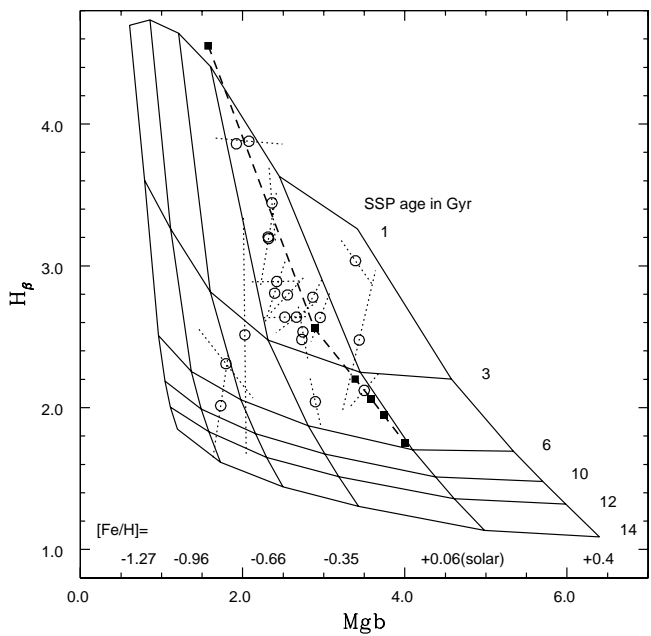}
\caption{The line index points from Figs.~\ref{fig:sfdgrid_m} and \ref{fig:sfd_exp_grid}are overlaid on a grid of simple stellar population models from BaSTI, with solar abundance points from MILES simple stellar populations added as solid squares.}
\label{fig:sfd_ssp_grid}
\end{figure}

\subsection{Simple stellar population models}

Figure~\ref{fig:sfd_ssp_grid} shows the data points from the previous two plots on a grid representing simple (single-age, single metallicity) stellar populations (SSPs), again from the BaSTI simulations.  The galaxy spectra are best fitted by SSPs with ages between 1 and 8~Gyr, with most being about 2 - 3~Gyr. 

It is important to test  the robustness of the derived parameters, and particularly the separation of emission and absorption components, to our specific choice of population synthesis models for predicting absorption line features, given the well-known differences in the predictions of Balmer line absorptions between different spectral synthesis libraries.  To investigate this, we did completely independent spectral fits using the MILES empirical spectral library of \cite{vazd10}.  This showed that the separation of spectral lines into emission and absorption components was largely unaffected by the choice of model used; however, the derived ages of the corresponding populations do change significantly depending on which model is used, with BaSTI predicting younger ages than MILES.  This systematic difference is the main uncertainty related to the choice of spectral model. We choose to continue to rely on BaSTI for our main age estimates, but since this is an important systematic, we have illustrated the differences in inferred BaSTI vs. MILES ages in Fig.~\ref{fig:sfd_ssp_grid}  Here the filled squares show MILES predictions of H$\beta$ and Mgb absorption strengths for SSPs with solar metallicity. This shows, for example, that the MILES 14\,Gyr SSP corresponds to a BaSTI 6\,Gyr model, though at the younger ages where the bulk of the SFD regions lie, the differences are much less marked, both in absolute and fractional terms. The most important conclusion from this comparison for the current work is that the separation of emission and absorption components is not significantly affected by the choice of stellar population model.

\section{Discussion}

\subsection{Comparison with theoretical predictions}

The ages we find can be compared to the theoretical predictions from simulations.  Some studies have concluded that bars overall tend to form in the early stages of galaxy evolution, and are subsequently stable and long-lived \citep{deba06,curi08}.  However, \cite{bour05} find that gas-rich galaxies have bars that are short-lived, and they predict bar durations of 1 - 2~Gyr in Sb - Sc type galaxies. \cite{atha13} use N-body simulations to investigate the formation of bars within disks, and find long timescales of several Gyr for stable bars to become established, a conclusion that is supported by \cite{mart17}.  A key finding of \cite{atha13} is that this timescale is longer for disks with higher gas-mass fractions\citep[see also][]{vill10}, corresponding to later-type classifications.  This is in agreement with the weak trend we find in Fig.~\ref{fig:sfdgrid_m}, discussed in section 4.3, where the latest type galaxies have the most recently established SFD regions, or indeed still have SF throughout their bars and inner disks as seen for NGC~2604.

In their study of a simulated Milky Way-like spiral galaxy, \cite{spin17} found the bar started to form at $z=$0.4, corresponding to a look-back time of 4.3~Gyr \citep{wrig06}, probably triggered by a minor merger; the bar stabilised and reached a maximum length at a look-back time of 1.3~Gyr ($z=$0.1), at which time it had substantially depleted the gas reservoir in a central region of diameter 4~kpc.  This depleted region persisted in their simulation to the present day, and strongly resembles our observed SFD regions in size and age.

\subsection{Comparison with other observational estimates of bar ages}

Many observational methods have been proposed in the literature to constrain bar ages and the evolution of bar properties with cosmological epoch.  \cite{joge04} studied the bar properties of a sample of high-redshift disk galaxies, finding that bars were in place early, and appeared to have long tifetimes, exceeding 2~Gyr. Other studies have inferred bar ages from detailed study of nearby barred galaxies. For example, \cite{gado05} measured the vertical velocity dispersion of the stellar component of strong bars, finding high dispersions ($\sim$100~km/s) that only developed in simulated galaxy bars after several Gyr.  In a subsequent paper \citep{gado06}, they used broad-band optical - near-IR colours to infer similarly old ages for the stellar populations that dominate bars.  In both of these studies, the age ranges are rather larger than the SF truncation ages found for most of the galaxies in the present sample, although at least for the study of \cite{gado06} it should be noted that constraints quoted are for the stellar components of the bar, which could be very different to the age of the bar itself. \cite{sanc11} studied two early type low-redshift barred galaxies, NGC\,1558 (type SAB0) and NGC\,1433 (SBab), concluding again that these bars are old and formed early in the lifetimes of their host galaxies. In a recent study, \cite{font17} analyse a large sample of barred galaxies with corotation radii derived from emission-line gas kinematics and find anomalous parameter correlations that they explain by invoking bar evolution over several Gyr, following a theoretical picture of gradual bar growth based on the models of \cite{mart17}. 

\subsection{A possible time sequence of bar-induced activity}

While there is still much work to be done on the details of the processes linked with bar activity in disk galaxies, it is possible to sketch out a possible time sequence of events following the first emergence of a bar, at least in terms of SF activity \citep[see also][]{mart97}.  The most probable initial response is enhancement of SF activity, as indicated by several studies, following the early work of \cite{heck80b}; see also \cite{hawa86,huan96,mart97,ague99,elli11,oh12}.  This short phase is likely to be marked by rapid transfer of gas to nuclear regions triggering strong nuclear SF, and SF along the bar itself, as is seen at the current epoch to be common in late type (SBc - SBm) barred galaxies (see e.g. NGC~2604 in Fig.~\ref{fig:barsx4}).
 
After this, the bar strengthens, lengthens, and suppresses SF over large radial range; this is the SFD phenomenon discussed in the present paper. This is a long-lived phase compared with the initial enhancement of SF, with Gyr timescales indicated by this analysis and the prevalence of this SF pattern in early and intermediate Hubble type barred galaxies.  During this phase, 
SF continues in the nucleus, and is induced in a ring around the bar ends, but these do not compensate for the complete suppression within the large area of the SFD, and so the overall effect is a reduction in the integrated SF rate of galaxy \citep{jame09}.

While this SFD phase is long-lived, the lack of ages $>$4~Gyr in the present study imply that the SFD phase does not last indefinitely. This raises the question of what happens after SFD phase: can the galaxy return to having a full SF disk, and potentially undergo multiple barred phases, or does the SF suppression propagate outwards to encompass the bar-end ring and outer disk, resulting in a red-sequence passive spiral or lenticular galaxy?  We plan to address this possibility in our future work.

\section{Conclusions}

The main conclusions of the present study are as follows:

\begin{itemize}
\item We confirm the SFD phenomenon, i.e. the strong suppression of SF in disk regions swept-out by strong galaxy bars using deep spectroscopy for a sample of 22 galaxies.
\item We are able to determine H$\beta$ absorption line EW values for disk regions of all but one of the observed galaxies. 
\item Absorption line strengths can be fitted by stellar populations with a constant SF rate prior to an abrupt and complete cessation of SF, with the epoch of this truncation event as the one free parameter in the models.
\item The truncation epoch varies significantly between different galaxies, over a range extending from 0.1~Gyr to more than 4~Gyr, and some evidence for a preferred age of about 1~Gyr.
\item Truncation ages derived from independent regions within the same galaxy are mutually consistent, given the expected measurement errors.
\item Stellar populations appear homogeneous within the SFD regions; we find no evidence for differences in spectroscopic properties on opposite sides of the nucleus within individual galaxies of this sample.
\item If the SF suppression is only 90 per cent effective, or if there is an equivalent level of radial mixing of young stars from outside the SFD region, the truncation epoch is shifted earlier, by a factor of approximately 2 in age.
\item Gradually-declining exponential SF histories fail to reconcile the population ages with the lack of current SF, supporting models with a specific epoch where SF is strongly truncated.
\item There is some indication of later and/or less complete truncation of SF in later-type barred galaxies, consistent with theoretical modelling of galaxies with higher gas-mass fractions, e.g. \cite{atha13}.
\end{itemize} 

Overall, these conclusions confirm and are fully consistent with those derived for a pilot study of 4 galaxies in \cite{jame16}. 

More work is needed to address several aspects of the processes discussed here. One remaining question concerns the physical process underlying the SFD phenomenon: is gas present in these regions, but stabilised against collapse, e.g. by bar-induced turbulence, or has it been removed completely through bar-driven torques?  Detailed mapping of gas densities within galaxy disks exhibiting the SFD phenomenon is required to answer this question.  

A second uncertainty concerns the continued development of the SFD galaxies.  Are there cases where the SF suppression continues to propagate outwards through the rings and outer disks, resulting in completely passive disk galaxies, without the involvement of external environmental or AGN-driven processes? What would the final galaxy state be as a result of such secular processes, and for example would the bars and rings seen in the SFD systems survive to the passive stage?  These are complex questions, requiring further theoretical input, but on the observational side, Integral Field Unit surveys yielding 2-dimensional stellar population maps of large samples of galaxies seem likely to play a strong role.
 
\section*{Acknowledgments}

We warmly thank the referee for useful suggestions which significantly improved the content and presentation of the paper. We are also happy to thank Martha Tabor for instrument setup and support at the Isaac Newton Telescope.
The Isaac Newton Telescope is operated on the island of La Palma by the Isaac Newton Group in the Spanish Observatorio del Roque de los Muchachos of the Instituto de Astrof\'isica de Canarias. 
This research has made use of the NASA/IPAC Extragalactic Database (NED) which is operated by the Jet Propulsion Laboratory, California Institute of Technology, under contract with the National Aeronautics and Space Administration.  
Funding for the SDSS and SDSS-II has been provided by the Alfred P. Sloan Foundation, the Participating Institutions, the National Science Foundation, the U.S. Department of Energy, the National Aeronautics and Space Administration, the Japanese Monbukagakusho, the Max Planck Society, and the Higher Education Funding Council for England. The SDSS Web Site is http://www.sdss.org/.
The SDSS is managed by the Astrophysical Research Consortium for the Participating Institutions. The Participating Institutions are the American Museum of Natural History, Astrophysical Institute Potsdam, University of Basel, University of Cambridge, Case Western Reserve University, University of Chicago, Drexel University, Fermilab, the Institute for Advanced Study, the Japan Participation Group, Johns Hopkins University, the Joint Institute for Nuclear Astrophysics, the Kavli Institute for Particle Astrophysics and Cosmology, the Korean Scientist Group, the Chinese Academy of Sciences (LAMOST), Los Alamos National Laboratory, the Max-Planck-Institute for Astronomy (MPIA), the Max-Planck-Institute for Astrophysics (MPA), New Mexico State University, Ohio State University, University of Pittsburgh, University of Portsmouth, Princeton University, the United States Naval Observatory, and the University of Washington.


\label{lastpage}

\end{document}